\documentclass{article}%
\usepackage{amssymb}
\usepackage{amsfonts}
\usepackage{amsmath}
\usepackage{graphicx}%
\providecommand{\U}[1]{\protect\rule{.1in}{.1in}}

\begin{document}

\title{\textbf{ A left-right $\mathbf{SU}(7)$ symmetric model with $\mathcal{D}%
-$parity cosmic strings}}
\author{Helder Ch\'{a}vez S.\thanks{E-mail: helderch@if.ufrj.br} \ and J. A. Martins
Sim\~{o}es\thanks{Corresponding author. E-mail address: simoes@if.ufrj.br
(J.A. Martins Sim\~{o}es)}\\Instituto de F\'{\i}sica, Universidade Federal de Rio de Janeiro}
\maketitle

\begin{abstract}
Cosmic strings with the property of $\mathcal{D}-$ parity symmetry are studied
in this paper. They are of a $\mathcal{Z}_{2}$ type of strings that could
appear in the spontaneous breaking of $SU(7)$ and would present extraordinary
properties in a background of ordinary and mirror neutrinos. Through the
special embedding of the left-right symmetry in $SU(7)$, with a minimal
content of Higgs fields, based on two singlets and two doublets, it is
possible to assure the topological stability of this type of cosmic strings.
In their presence we could have a neutral flavor changing interaction between
ordinary and mirror neutrinos as well as the formation of superconducting
currents in the form of zero modes of neutrino mirrors that would show
interesting effects.

\end{abstract}

\section{\textbf{Introduction}}

The cosmological scenario generally accepted today is that the universe, in
its cooling process, has suffered a sequence of phase transitions in which
symmetries were spontaneously broken until arriving to the symmetry of nature
as we presently observe: $SU(3)_{C}\otimes U(1)_{\text{e.m}}$\cite{1}.In
consequence, according to the Kibble mechanism \cite{2} the formation of
extended topological objects like cosmic strings could have arisen if the
topology of the vacuum manifold of the symmetry is nontrivial. Topological
defects are also real objects in low energy physics of the condensed matter.
Some well-known examples are flux tubes in superconductors and vortices in
superfluid helium-4 \cite{3}.

There are two forms of classifying cosmic strings: the first one uses the
Wilson-line integral at infinite radius $U\left(  \theta\right)  =P\exp\left[
\int_{0}^{\theta}\mathbf{A.}d\mathbf{l}\right]  ,$ where $P$ represents the
path ordering of the exponential. This generates the condensate winding at
spatial infinity $\left\langle \phi\left(  \theta\right)  \right\rangle
=U\left(  \theta\right)  \left\langle \phi\left(  0\right)  \right\rangle $,
where $\left\langle \phi\right\rangle $ is the vacuum expectation value of the
Higgs field producing the breaking $G\,\underrightarrow{\left\langle
\phi\right\rangle }\,H.$ This is the case for $U\left(  2\pi\right)  =h\in H,$
being $H$ the little group of the Higgs field $\left\langle \phi\left(
0\right)  \right\rangle $. Thus, the cosmic strings that can be formed in this
breaking are specified by the possible values of $U\left(  2\pi\right)  -$ the
elements of $H.$ The other form of classifying cosmic strings is to specify
the topological class of a string, such that the vacuum manifold $G/H$ has a
nontrivial fundamental group, i.e., the elements of the first homotopy group
$\pi_{1}\left(  G/H\right)  $ are nontrivial. In general, $m-$dimensional
defects in a $d-$dimensional medium are classified by the homotopy group
$\pi_{n}\left(  G/H\right)  $ where\cite{4} $n=d-m-1,$ such that, when
$n=0,1,2$ \ the objects formed are domain walls, cosmic strings and magnetic
monopoles, respectively.

A cosmic strings model based in the product group $SO(10)\otimes SO(10)%
\acute{}%
\subset SO(20)$ GUT with $SO(10)$ being the symmetry of ordinary matter and
$SO(10)%
\acute{}%
$ \ describing the mirror matter was constructed by Schwarz \cite{5}. This
kind of objects have the extraordinary property of transforming an ordinary
particle in a mirror particle when this particle gives a turn around the
string. This type of cosmic strings are considered as an Alice string
\cite{6}. A model of mirror cosmic strings as possible sources of ultrahigh
energy neutrinos was constructed by Berezinsky et al.\cite{7}. Some models of
left-right cosmic strings \cite{8}\cite{9} as well as $B-L$ cosmic strings
also exists in the literature \cite{10}. Although explicit supersymmetric
models of cosmic strings have been built \cite{11}, explicit models of cosmic
strings in $SU(N>5)$ GUT%
\'{}%
s with mirror matter have not been considered in the literature. Some of the
difficulties of this class of models were to reproduce the well-known
phenomenology of low energy. But it is also possible that this kind of objects
have been formed in the TeV scale or in the electro-weak scale, with the
possibility of having experimentally sizeable effects.

The connection between left-right symmetric models and cosmic strings is very
appealing but finds a fundamental phenomenological difficulty: the Higgs
sector that breaks the parity symmetry has a large number of unknown
fundamental parameters and we have no direct connection with the cosmic
strings scales.

Recently we have studied \cite{14,15} a new mirror left-right symmetric model.
In this model we have shown that $SU(2)_{L}\otimes SU(2)_{R}\otimes
U(1)_{B-L}$ can be broken to the standard model group with a Higgs sector
containing only two doublets and two singlets. In this model parity is broken
by the $\mathcal{D-}$ parity mechanism and an $SU(7)$ GUT model is proposed.

In the present paper we have constructed a cosmic strings model of the type
$\mathcal{Z}_{2}$, and have called it a $\mathcal{D-}$ parity cosmic strings,
based in a $SU(7)$ GUT with a especial embedding in order to incorporate the
minimal left-right symmetry as a subgroup and include mirror matter. A careful
election of the Higgs fields is necessary to get the topological stability of
this cosmic strings until our days.

We have also the possibility of $B-L$ cosmic strings in our model, but the
presence of $SU(2)_{R}$ would desestabilize them \cite{8}. This type of
strings are produced when a factor $U(1)_{B-L}$ (that contain a $\mathcal{Z}%
_{2}$ discrete symmetry which can be left unbroken down to low energies) is
broken by a Higgs scalar in a complex representation of $G$.

The $P-$parity spontaneously breaking will happen with the break of the
left-right components in the low energy stage governed by the symmetry
$SU(3)_{C}\otimes SU(2)_{L}\otimes SU(2)_{R}\otimes U(1)_{B-L}$ \cite{12}%
\cite{13}\cite{14}.

Our paper is organized as follow: after making some general comments in the
introduction, in the section 2 we analyze the breakdown of $SU(7)$ and the
consequent appearance in our model of $\mathcal{D-}$ parity\textbf{ }cosmic
strings. In section 3 we present the question of magnetic monopoles in the
phase transitions. The mechanism of generation of superconducting currents
with neutrino mirrors as well as the flavor changing between ordinary and
mirror neutrinos is approached in section 4. In section 5, final comments and
conclusions are provided.

\section{\textbf{Breakdown of SU(7) and} $\mathcal{D-}$ \textbf{parity cosmic
strings}}

The spontaneous breaking\ of $SU(N)$ gauge theories usually can be made
through the fundamental representation $\mathbf{N}$ or with multiplets
corresponding to the direct product $(\mathbf{N}^{\mathbf{2}}\mathbf{-1}%
)\otimes(\mathbf{N}^{\mathbf{2}}\mathbf{-1})$. However, to find Higgs
multiplets producing $\mathcal{Z}_{2}$ cosmic strings in a semisimple gauge
group $G$ it is necessary to observe that the chiral fermions must be placed
in a irreducible fundamental representation. Then it is necessary to look for
the product $\mathbf{N}\otimes\mathbf{N}$ for the symmetrical component of
higher weight giving masses to the fermions\cite{21}. This is the habitual
procedure for $G=SO(10),E_{6},E_{7},E_{8}.$ Nevertheless, in $SU(N)$ grand
unified gauge theories the chiral fermions are placed in combinations of
representations in such a way that they eliminate the anomalies. For $SU(7),$
our election, free of anomalies, is $\mathbf{\{7\}\oplus\{21}^{\ast
}\mathbf{\}\oplus\{35\}.}$

An interesting class of cosmic strings arises from the following breaking
chain of $SU(7)$
\begin{gather}
SU(7)\ \underrightarrow{S_{M}}\ SU(5)\otimes SU(2)_{R}\otimes\mathcal{Z}%
_{2}\ \underrightarrow{S_{D}}\ \nonumber\\
SU(3)_{C}\otimes SU(2)_{L}\otimes SU(2)_{R}\otimes U(1)_{B-L}\otimes
\mathcal{Z}_{2}\,\underrightarrow{\chi_{R}}\ \nonumber\\
\ SU(3)_{C}\otimes SU(2)_{L}\otimes U(1)_{Y}\otimes\mathcal{Z}_{2}%
\ \underrightarrow{\chi_{L}}\ SU(3)_{C}\otimes U(1)_{e.m}\otimes
\mathcal{Z}_{2}. \tag{1}%
\end{gather}
The matter content is included in the decomposition of the representations of
the symmetry breaking Higgs fields and the fundamental fermions multiplet
under the $SU(5)\otimes SU(2)_{R}\otimes U(1)_{X}$ maximal
subgroup\footnote{Our notation is: $\left\{  \quad\right\}  $ for $SU(7)$,
$\left[  \quad\right]  $ for the components under $SU(5)\otimes SU(2)_{R}%
\otimes U(1)_{X}$ and $\left(  \quad\right)  $ for $SU(3)_{C}\otimes
SU(2)_{L}\otimes SU(2)_{R}\otimes U(1)_{B-L}$ decompositions. It should be
noticed that \{63\} it is not an irreducible representation of $SU(7)$.} of
$SU(7)$:%
\begin{gather}
\{\mathbf{63\}=\{7\}\oplus\{21}^{\ast}\mathbf{\}\oplus\{35\},}\tag{2}\\
\{\mathbf{21}^{\ast}\}=\mathbf{[1,1,}10]\oplus\lbrack\mathbf{5}^{\ast
}\mathbf{,2},3]\oplus\mathbf{[10}^{\ast},\mathbf{1},-4],\tag{3}\\
\mathbf{\{48\}=[24,1,}0\mathbf{]\oplus...} \tag{4}%
\end{gather}
where we have indicated only those pieces that acquire vacuum expectation
values for $\mathbf{\{21}^{\ast}\mathbf{\}}$ and $\mathbf{\{48\}}$.

Our election for the Higgs multiplets is $S_{M}\sim\{\mathbf{21}^{\ast
}\}\oplus\mathbf{\{1\}}\supset\mathbf{[1,1,}10]\oplus\lbrack\mathbf{1,1,}%
0\mathbf{]},$ i.e., $S_{M}$ is some linear combination of Higgs fields in the
$\{\mathbf{21}^{\ast}\}$ and $\mathbf{\{1\}}$ representations along with their
coupling strengths. The first component $S_{M}\left(  \mathbf{21}\right)
\sim\{\mathbf{21}^{\ast}\}\supset\mathbf{[1,1,}10]\sim(\mathbf{1,1,1,}0)$ it
should be responsible for producing cosmic strings.\ We clarify our election
of the Higgs field breaking $SU(7)$. In reality, from the product of the
fundamental representations $\mathbf{\{7\}\otimes\{7\}=\{21\}}_{A}%
\oplus\{\mathbf{28}\}_{S}$ we can see that $\{\mathbf{28}\}=[\mathbf{15,1,}%
4]\oplus\lbrack\mathbf{5,2,}-3]\oplus\lbrack\mathbf{1,3,}-10]$ doesn't contain
any component that can break $SU(7)$ leaving invariant $SU(5)\otimes
SU(2)_{R}$ in the next stage, as it is our desire.\ A different situation
happens for $\mathbf{\{21\}}_{A}$ that contain the appropriate piece
$\mathbf{[1,1,-}10]$ in order to construct cosmic strings $\mathcal{Z}_{2}$
with the surprising property of changing flavors of ordinary and mirrors
neutrinos. However, as it was observed in reference \cite{25}, in unified
theories it is also possible to generate fermion masses using antisymmetric
representations. Although in our model neutrinos don't receive masses at the
tree level from $\mathbf{\{21\}}_{A}$ at the GUT scale, it is vital to
generate radiative neutrino masses \cite{15}. For this reason, we have done
the election of antisymmetric components of $\mathbf{N\otimes N}$ that would
also produce different types of cosmic strings $\mathcal{Z}_{2}$.

The second term $S_{M}\left(  \mathbf{1}\right)  \sim\mathbf{\{1\}\supset
\,}[\mathbf{1,1,}0\mathbf{]}$ could generate superheavy mass to some neutrinos
as well as driving the inflation scenario, as we will see below. The following
decompositions are necessary%
\begin{gather}
\mathbf{\{35\}=\,}[\mathbf{10}^{\ast}\mathbf{,1,}6]\oplus\lbrack
\mathbf{5,1,}-8]\oplus\lbrack\mathbf{10,2,}-1],\nonumber\\
\mathbf{\{48\}=[1,1,}0\mathbf{]\oplus}[\mathbf{1,3},0]\oplus\lbrack
\mathbf{24,1,}0]\oplus\left[  \mathbf{5,2,}7\right]  \oplus\left[
\mathbf{5}^{\ast}\mathbf{,2,-}7\right]  ,\nonumber\\
\{\mathbf{224}\}=\mathbf{[40\mathbf{,1,}}4\mathbf{]\oplus\lbrack
24\mathbf{,1,}}-10\mathbf{]\oplus\lbrack45\mathbf{,2,}}-3\mathbf{]\oplus
\lbrack10}^{\ast}\mathbf{\mathbf{,2,}}11\mathbf{]\oplus\lbrack5\mathbf{,2,}%
}-3\mathbf{]}\nonumber\\
\mathbf{\oplus\lbrack10\mathbf{,1,}}4\mathbf{]\oplus\lbrack10}%
,\mathbf{3\mathbf{,}}4\mathbf{],} \tag{5}%
\end{gather}
The next important fields are $S_{D}\sim\mathbf{\{48\}\supset\lbrack
24,1,}0\mathbf{]\supset(1,1,1,}0)\mathbf{,}$ $\chi_{R}\sim
\mathbf{\{35\}\supset}$ $[\mathbf{10,2},-1]\supset\mathbf{(1,1,2,}1)$ and also
$\chi_{L}\sim\mathbf{\{224\}\supset\,}[\mathbf{10,1},-4]\supset$
$(\mathbf{1,2,1},1)$ which are invariant under $2\pi$ rotations because they
are not spinorial representations$.$

The fermions content is deployed \cite{15} in the
representations$\ \mathbf{\{1\},\{7\},\{21\}}$ and $\mathbf{\{35\}}$ with the
anomaly free combinations $\mathbf{\{1\}\oplus\{7\}\oplus\{21}^{\ast
}\mathbf{\}\oplus\{35\}}$. These selected Higgs multiplets can give masses to
all the fermions as it can be directly verified from the tensorial products
\cite{16}:%
\begin{gather}
\mathbf{\{7}^{\ast}\mathbf{\}\otimes\{7\}=\{1\}\oplus\{48\},}\nonumber\\
\mathbf{\{7\}\otimes\{7\}=\{21\}}_{A}\mathbf{\oplus\{28\}}_{S},\nonumber\\
\mathbf{\{7\}\otimes\{35\}=\,}\{\mathbf{35}^{\ast}\}\oplus\mathbf{\{210\},}%
\nonumber\\
\mathbf{\{7}^{\ast}\mathbf{\}\otimes\{35\}=\{224\}\oplus\{21\},}\tag{6}\\
\mathbf{\{21\}\otimes\{35\}=\{224\}\oplus\{21}^{\ast}\mathbf{\}\oplus
...,}\nonumber\\
\mathbf{\{21}^{\ast}\mathbf{\}\otimes\{21\}=\{1\}\oplus\{48\}\oplus
...,}\nonumber\\
\mathbf{\{21\}\otimes\{21\}=(\{196\}\oplus\{35\})}_{S}\mathbf{\oplus
\{210\}}_{A}\mathbf{,}\nonumber\\
\mathbf{\{35}^{\mathbf{\ast}}\mathbf{\}}\otimes\mathbf{\{35\}=\{1\}\oplus
\{48\}\oplus...}\nonumber
\end{gather}
The remaining particles that didn't obtain their masses at the tree level,
will obtain their masses from radioactive corrections.

The $\mathcal{Z}_{2}$ symmetry that appears in (1) is the discrete remanent of
a broken $U(1)_{X}$, which can be identified as $\mathcal{D-}$ parity symmetry
\cite{13} as is evident from the next stage of symmetry breaking.

Now we pass to describe the breaking chain (1). If one begins with a simply
connected gauge group $G,$ strings will not arise in a phase transition in
which $G$ is spontaneously broken. This is the case for $SU(7)$ as in any GUT
based in $SU(N)$ or $Spin(N);$ strings will not arise at the first stage of
symmetry breaking \cite{17}. Then, the breakdown of $SU(7)$ to its maximal
sub-group $SU(5)\otimes SU(2)\otimes U(1)_{X}$ will not produce strings. The
previous $SU(2)$ factor is assumed to be the right sector, i.e. we assume
$SU(2)_{R}$. This is possible because the generators of $SU(2)_{R}$ are
included in $SO(14)$ and $SU(7)$ is naturally embedded in $SO(14)$.

However, $SU(5)\otimes SU(2)_{R}$ is connected and by virtue of a well-known
property of the homotopy groups\cite{4} $\pi_{i-1}SU(n)=\pi_{i+1}%
SU(n),i\leqslant2m\leqslant n$. So we have $\pi_{0}(SU(5))=0$ and $\pi
_{0}(SU(2))=0,$ and then
\begin{equation}
\pi_{1}\left(  \frac{SU(7)}{SU(5)\otimes SU(2)_{R}\otimes\mathcal{Z}_{2}%
}\right)  =\pi_{0}(SU(5)\otimes SU(2)_{R}\otimes\mathcal{Z}_{2})=\mathcal{Z}%
_{2}.\tag{7}%
\end{equation}
The kind cosmic strings formed in this stage of breaking symmetry has energy
per unit length $\sim$ $\left\langle S_{M}\left(  \mathbf{21}\right)
\right\rangle ^{2}$ and could be called $D-$\textit{parity cosmic strings }for
arguments that we will be giving soon. According to the products given in (6),
the following relevant couplings conserving the $U(1)_{X}$ charge are
possible:%
\begin{gather}
\mathbf{\{21}^{\ast}\mathbf{\}\otimes\{1\}\otimes\{21}{}_{H}\}\supset
\overline{\nu_{eR}}N_{EL}\widetilde{S_{M}}\left(  \mathbf{21}\right)
,\nonumber\\
\mathbf{\{1\}\otimes\{1\}\otimes\{1}_{H}\mathbf{\}}\sim\overline{N_{EL}^{C}%
}N_{EL}S_{M}\left(  \mathbf{1}\right)  ,\nonumber\\
\mathbf{\{7}^{\ast}\}\otimes\mathbf{\{7\}\otimes\{1\}}_{H}\supset\overline
{\nu_{\mu R}^{C}}\nu_{\mu R}S_{M}\left(  \mathbf{1}\right)  ,\overline
{N_{ML}^{C}}N_{ML}S_{M}\left(  \mathbf{1}\right)  ,\tag{8}\\
\mathbf{\{7}\}\otimes\mathbf{\{7\}\otimes\{21}_{H}^{\ast}\mathbf{\}}%
\supset\overline{\nu_{\mu R}}N_{ML}S_{M}\left(  \mathbf{21}\right)  \nonumber
\end{gather}
Similar couplings arise for the multiplets in $\mathbf{64}^{\ast
}=\mathbf{\{1\}\oplus\{7}^{\ast}\mathbf{\}}\oplus\mathbf{\{21\}\oplus
\{35}^{\ast}\mathbf{\}}$, where we can accommodate the lepton $\tau$ family, a
new fourth family of "ordinary" leptons $l_{\theta L}^{C}=\dbinom{\nu_{\theta
L}}{\theta_{L}}^{C}$ $\subset$ $\mathbf{\{7}^{\ast}\mathbf{\}}$, its quarks
$q_{L}^{C}=\dbinom{o_{L}}{a_{L}}^{C}\subset\mathbf{\{35\}}^{\ast}$ and its
respective mirror partners \cite{15}.

In the following stage for the breaking chain (1) it is possible that $B-L$
cosmic strings could be formed since%
\begin{equation}
\pi_{1}\left(  \frac{SU(3)_{C}\otimes SU(2)_{L}\otimes SU(2)_{R}\otimes
U(1)_{B-L}}{SU(3)_{C}\otimes SU(2)_{L}\otimes U(1)_{Y}}\right)  =\mathcal{Z}%
_{2}.\tag{9}%
\end{equation}
Thus, as $\mathcal{Z}_{2}$ it is not already contained in a continuous
connected invariance group $SU(3)_{C}\otimes SU(2)_{L}\otimes U(1)_{Y}$ of the
vacuum, $B-L$ cosmic strings of energy per unit length $\sim$ $\left\langle
\chi_{R}\right\rangle ^{2}$ can appear in this stage. However, it was showed
in \cite{8} that this $\mathcal{Z}_{2}$ as expected from the breakdown of
$U(1)_{B-L}$ group, by itself does not persist due to the presence of the
$SU(2)_{R}.$ Thus, the $B-L$ cosmic strings appearing in this stage will be
unstable and decaying quickly. In the phenomenological context, it is also
important to notice that $\mathcal{P}-$ parity is spontaneously broken along
with the group $SU(2)_{R}$. Some low energy phenomenologic aspects of
$SU(3)_{C}\otimes SU(2)_{L}\otimes SU(2)_{R}\otimes U(1)_{B-L}$ with mirror
matter were studied in \cite{15}\cite{18}.

\section{\textbf{Monopoles in the phase transitions}}

In this section we look for the formation of other topological defects, such
as monopoles, in the breaking chain (1). As it was recognized in the
literature, the symmetry of matter parity $\mathcal{Z}_{2}$ is important in
order to preserve large values for the proton lifetime, and also because it
guarantees the topological stability of cosmic strings. Let us use the Kibble
mechanism \cite{2} based in homotopy theory to find monopoles. Thus $\pi
_{2}\left(  \frac{SU(7)}{SU(5)\otimes SU(2)_{R}\otimes\mathcal{Z}_{2}}\right)
=\pi_{1}\left(  SU(5)\otimes SU(2)_{R}\otimes\mathcal{Z}_{2}\right)
=\mathcal{I}$\ is trivial. According to the Kibble mechanism \cite{2}
topological monopoles are not formed during the first phase transition in (1).
Let us also notice the relations%
\begin{align}
&  \pi_{2}\left(  \frac{SU(7)}{SU(3)_{C}\otimes SU(2)_{L}\otimes
SU(2)_{R}\otimes U(1)_{B-L}\otimes\mathcal{Z}_{2}}\right)  \nonumber\\
&  =\pi_{1}\left(  SU(3)_{C}\otimes SU(2)_{L}\otimes SU(2)_{R}\otimes
U(1)_{B-L}\otimes\mathcal{Z}_{2}\right)  =\mathcal{Z},\tag{10}%
\end{align}%
\begin{align}
&  \pi_{2}\left(  \frac{SU(7)}{SU(3)_{C}\otimes SU(2)_{L}\otimes
U(1)_{Y}\otimes\mathcal{Z}_{2}}\right)  \nonumber\\
&  =\pi_{1}\left(  SU(3)_{C}\otimes SU(2)_{L}\otimes U(1)_{Y}\otimes
\mathcal{Z}_{2}\right)  =\mathcal{Z},\tag{11}%
\end{align}%
\begin{equation}
\pi_{2}\left(  \frac{SU(7)}{SU(3)_{C}\otimes U(1)_{e.m}\otimes\mathcal{Z}_{2}%
}\right)  =\pi_{1}(SU(3)_{C}\otimes U(1)_{e.m}\otimes\mathcal{Z}%
_{2})=\mathcal{Z}.\tag{12}%
\end{equation}
Thus, starting from the second phase transition in the breaking chain (1),
topological magnetic monopoles were formed which will be topologically stable
until low energies. These objects, if they are present until our days, would
be dangerous for the universe because they would dominate the energy density
quickly. Consequently, it is necessary to appeal to inflation to dilute them.
This is possible if the singlet $SU(7),$ $\mathbf{\{1\}}$ is assumed to be the
responsible to generate the inflation scenario. We assume a coupling between
$S_{D}\sim\mathbf{\{48\}}$ producing monopoles, and $\mathbf{\{1\}}$ driving
the inflation by means of a hybrid inflation potential, for example as given
in \cite{19}. This mechanism, together with the monopole-antimonopole pair
nucleation could dilute necklaces cosmic strings \cite{20} possibly formed in
this phase transition. To avoid that cosmic strings are thrown away or
dissociated as consequence of the inflation, we suppose some discrete symmetry
to avoid a coupling between the $\mathbf{\{1}_{H}\mathbf{\}}\sim
S_{M}\mathbf{(1)}$ and the $\mathbf{\{21}_{H}^{\ast}\mathbf{\}}\sim
S_{M}(\mathbf{21)}$.

\section{\textbf{Neutrino effects in} $\mathcal{D}-$ \textbf{parity}
\textbf{cosmic strings}}

We begin with the first phase transition $SU(7)\ \underrightarrow
{\widetilde{S_{M}}}\ SU(5)\otimes SU(2)_{R}\otimes\mathcal{Z}_{2}$. The type
of cosmic strings taking place in this phase transition is $\mathcal{Z}_{2}$
\cite{21}. The only supermassive fermions in this stage are $N_{EL}$;
\textbf{$\nu_{\tau R}$} and it%
\'{}%
s mirror partner $N_{\Upsilon L}$ and $\nu_{\mu R}$ with it%
\'{}%
s mirror partner $N_{ML}$. They obtain Majorana masses of order $10^{16}GeV$
\ through the component $\mathbf{\{1}_{H}\mathbf{\}}\sim S_{M}\mathbf{(1).}$

A mixing of the type $\overline{\nu_{eR}}N_{EL}\widetilde{S_{M}}\left(
\mathbf{21}\right)  \subset\mathbf{\{21}^{\ast}\mathbf{\}\otimes
\{1\}\otimes\{21}{}_{H}\}$ has interesting effects. The cosmic strings
solutions are given by the classical configurations $S_{M}^{clas}\left(
\mathbf{21}\right)  =f\left(  r\right)  e^{i\theta}S_{M}^{(0)}\left(
\infty\right)  $ and $A_{\theta}^{clas}=\frac{2a\left(  r\right)  }{gr}%
\delta_{\theta}^{\mu}$; where we are assuming the winding number $n=1$ and the
vacuum expectation value is $\left\langle S_{M}^{(0)}\left(  \infty\right)
\right\rangle =\upsilon_{M}/\sqrt{2}$ \cite{1}. Thus, in this case a
supermassive mirror electron neutrino $N_{EL}$ coming closer from the space
infinity to the anti-cosmic string, where $a(r),$ $f(r)\rightarrow1,$ after
giving a complete turn of $2\pi$ around of the anti-cosmic string becomes in
$\nu_{eR}.$ An analogous situation will be present between $\nu_{\tau R}$ and
their mirror partner $N_{\Upsilon L}.$

The formation of zero modes is also possible if we add the quantum
fluctuations to the classical configurations of the Higgs and gauge fields of
the string: $S_{M}\left(  \mathbf{21}\right)  =S_{M}^{clas}\left(
\mathbf{21}\right)  +\widehat{S_{M}}\left(  \mathbf{21}\right)  $, $A_{\mu
}=A_{\mu}^{clas}+\widehat{A}_{\mu}$ in an analogous way as to the capture of
an electron by a nucleus with the emission of a photon. In this sense, it is
expected the capture of $N_{EL}$ by the anti-string and the subsequent
formation of zero mode of $\nu_{eR}$ with the emission of the scalar Higgs or
the vectorial boson $A_{\mu}$ that form the anti-string. The inverse process
is also possible. If the necessary kinematic considerations are allowed, we
can have the capture of an $\nu_{eR}$ and the formation of neutral currents
with the mirror electron neutrino $N_{EL}$ and the emission of bosons from the
string. One should not forget that at this stage these currents are massless
because in the string we have $a(r),\,f\left(  r\right)  \rightarrow0$ as
$r\rightarrow0.$

A similar situation will take place through the coupling $\mathbf{\{7}%
\}\otimes\mathbf{\{7\}\otimes\{21}_{H}^{\ast}\mathbf{\}}\supset\overline
{\nu_{\mu R}}N_{ML}S_{M}\left(  \mathbf{21}\right)  $ where a mirror muon
neutrino\ $N_{ML}$ giving a complete turn of $2\pi$ around of the cosmic
string becomes a $\nu_{\mu R}$ with the possibility of forming massless zero
modes by means of the same mechanism as described before. Thus, it is possible
to generate superconducting currents with mirror neutrinos. Similar ideas were
placed by one of the authors in a model of superconducting cosmic strings
$SO(10)$ in order to explain UHECR \cite{22}. Processes of flavor changing
neutral currents in cosmic strings and domain walls were also analyzed in
\cite{23}. The extraordinary consequence of our model is that in the presence
of this type of cosmic string, flavor changes of neutrinos would take place
through $N_{EL}\leftrightarrow\nu_{eR},N_{\Upsilon L}\leftrightarrow\nu_{\tau
R}$, $N_{ML}\leftrightarrow\nu_{\mu R},N_{\Theta L}\leftrightarrow\nu_{\theta
R}$. This is the reason why we call these strings $\mathcal{D}-$ parity cosmic
strings. It is also possible that this type of topological defects could hide
mirror neutrinos until today in the form of massless zero modes.

\section{\textbf{Comments and conclusions }}

We have built a model of $\mathcal{Z}_{2}$ cosmic strings which we have called
$\mathcal{D}-$ parity cosmic strings in virtue to their extraordinary property
of changing flavor between ordinary neutrinos and mirrors neutrinos, in the
GUT scale, when one of them gives a complete turn around the string. Our type
of string is not an Alice string in which a particle is transformed in its
anti-particle. This extraordinary property of this type string is a
consequence of the mixing between ordinary neutrinos and their mirror partners
through the field of a Higgs particle in the string and of the magnetic flow
inside the string that makes a rotation of the fermion field approaching to it
from very far.

A more realistic model of $\mathcal{D-}$parity cosmic strings could arise from
a breaking chain of the type%
\begin{gather}
G\rightarrow\widetilde{H}\otimes\mathcal{Z}_{2}\rightarrow H\otimes
\mathcal{D}\otimes\mathcal{Z}_{2}\rightarrow SU(3)_{C}\otimes SU(2)_{L}\otimes
SU(2)_{R}\otimes U(1)_{B-L}\otimes\mathcal{Z}_{2}\nonumber\\
\rightarrow SU(3)_{C}\otimes SU(2)_{L}\otimes U(1)_{Y}\otimes\mathcal{Z}%
_{2}\rightarrow SU(3)_{C}\otimes U(1)_{e.m}\otimes\mathcal{Z}_{2},\tag{13}%
\end{gather}
where $G=SO(14),$ $\widetilde{H}=SU(7)$ and $H=SU(5)\otimes SU(2)_{R}.$ In
reality, the first phase transition it can be produced for an Higgs
in\footnote{Here $\mathbf{R-}$representations correspond to $SO(14).$}
$\mathbf{1716}_{S}\,\mathbf{\subset64\otimes64}$ through of it%
\'{}%
s singlet component that leaves invariant the factor $SU(7)$. The second phase
transition could be produced for $\mathbf{91\supset\{1\}\supset\lbrack
1,1,}0\mathbf{]}$ which is even under $\mathcal{Z}_{2}$ and the third phase
transition by other $\mathbf{91\supset\{21\}\supset\,}[\mathbf{1,1,}-10]$
which is odd under $\mathcal{D}-$parity but is even under $\mathcal{Z}_{2}.$
Thus, the Left-Right hierarchy is induced of natural way in our breaking
chains (13). In the presence of mirror matter this type of cosmic strings
could have important effects \cite{24}.

Superheavy neutrinos mirror could also be the source of UHECR through the
mechanism of generation of superconducting currents described here. In this
context, if they were captured at the beginning of the friction period, an
estimative of the vortons density indicates that it would be more relevant for
UHECR in this period than in the scaling regime \cite{22}.

\begin{description}
\item[\textbf{Acknowledgment}] We thank Marco A. C. Kneipp for useful comments
of this paper. Helder Ch\'{a}vez S. thanks especially to FAPERJ for the
financial support.
\end{description}


\begin{thebibliography}{99}
\bibitem {24}At the moment we are studying cosmic strings arising from the
phase transitions (12) and their effects in a background of matter mirror.
\end{thebibliography}
\end{document}